\documentstyle[11pt,aas2pp4,epsfig]{article}
 
\newcommand{\hii}{H\,{\sc ii}}
\newcommand{\caii}{Ca\,{\sc ii}}
\newcommand{\ha}{H$\alpha$}
\newcommand{\msun}{M$_\odot$}
\newcommand{\zsun}{Z$_\odot$}
\newcommand{\cosp}{CO$_{\rm sp}$}  
\newcommand{\mdot}{$\dot{\rm M}$}

\begin{document}

\title{Usage of Red supergiant spectral features as age indicators 
       in starburst regions}

\author{Y.D. Mayya\altaffilmark{1}}
\affil{Instituto Nacional de Astrofisica, \'Optica y Electr\'onica,
       Apdo Postal 51 y 216, 72000 Puebla, Pue., M\'EXICO}
\affil{ydm@inaoep.mx}
\authoremail{ydm@inaoep.mx}
\altaffiltext{1}{Fellow of the {\it Programa Internacional de
Astrofisica Avanzada ``Guillermo Haro''}}
\affil{(To appear in ApJ lett.)}

\begin{abstract}

We investigate techniques that can be used to determine 
ages of starburst regions containing populations beyond 
their early nebular phase. In particular, we study the 
strength of the \caii\ triplet ($\lambda\lambda$ 8498, 8542, 8662 \AA) 
and the CO index (2.31--2.40 $\mu$m band) using synthetic models as 
the starburst evolves. For an instantaneous burst of star formation 
both of these absorption features remain strongest between 7--14~Myr 
corresponding to the red supergiant population. The detailed 
evolutionary behavior of the starburst is strongly metallicity 
dependent. Low metallicity starburst models successfully 
reproduce the distribution of equivalent widths of \caii\ triplet with 
age in Large Magellanic Cloud clusters. The clusters in the red 
supergiant phase strongly favor the stellar evolutionary models 
incorporating mass-loss rates higher than the standard values. 
We suggest usage of diagrams involving \caii\ triplet equivalent width, 
CO index and nebular recombination lines to infer the history as 
well as age of starburst regions.

\end{abstract}

\keywords{Galaxies: Magellanic Clouds --- galaxies: starburst --- 
            stars: supergiants}

\section{Introduction}

Detection of \caii\ triplet (CaT) lines in a circumnuclear \hii\
region of the starburst galaxy NGC\,3310 by Terlevich et al. (1990)
offered a new tool to study aging starburst systems. The CaT
lines originate in the atmospheres of cool stars with the absorption 
being the strongest in red supergiants (RSGs). It is known from the stellar 
evolutionary models at solar metallicity (\zsun$=$0.02) that O stars 
in the mass range 15--25\msun\ evolve into RSGs in 7--14~Myr, and hence the
detection of the CaT lines in starburst systems is often used to
infer the presence of a stellar population of around 10~Myr age. 
However, the CaT line equivalent widths of cool giants are 
non-negligible and hence the detection of the triplet lines is not an
unambiguous signature of the presence of RSGs. 

Bica, Alloin \& Santos (1990) provided a solution to this problem 
by obtaining the CaT equivalent widths of blue clusters 
with well determined turn-off ages in the Large Magellanic Cloud (LMC). 
They found that only clusters with an age $\sim$10~Myr have CaT 
equivalent widths greater than 6\AA; the older clusters having
values around 4\AA. However, their observed relation has only limited use 
in estimating the age of starburst regions due to the fact that the 
evolutionary history of massive stars, and hence the presence of RSGs, 
is a strong function of metallicity (Maeder 1991), and that 
starbursts are found in a wide range of metallic environments.
Thus a quantitative interpretation of the observed equivalent widths in 
starbursts requires studies of the evolutionary behavior of starbursts 
at different metallicities. We carry out such a study here
with special emphasis on reproducing the observed relations of 
Bica et al. (1990). We also study the evolutionary behavior of the CO index,
which is also sensitive to the population of cool stars, to investigate its 
use as an age indicator in conjunction with the CaT equivalent 
widths. As this is one of the first attempts in synthesizing these spectral 
features (see Leitherer et al. 1996 for a list of on-going attempts)
we mainly focus on understanding the basic evolutionary properties and
its metallicity dependence. 
The synthesis technique and the stellar data used are described in Sec.~2.
Evolutionary results are presented in Sec.~3 and are discussed from 
the context of observational data of Bica et al. (1990) for LMC
clusters in Sec.~4. The results are discussed from the point of view
of age estimations of starburst regions in Sec.~5.

\section{Synthesis model and Stellar Data}

We use the conventional evolutionary population synthesis technique
to compute the starburst related quantities.
The synthesis code we use in the computation is described in detail
in Mayya~(1995) and the results are compared with other existing
codes in Charlot~(1996). The major change in the present code is the
implementation of the isochrone synthesis technique (Charlot \& Bruzual 1991) 
which has the effect of reducing the noise on the computed quantities. The
equivalent width of the CaT
absorption lines is computed for
an instantaneous burst (IB) of star formation using the formula,
$$
{\rm EW}_{{\rm CaT}}({\rm imf},t,Z) = 
       {{\int\limits_{m_l}^{m_u}L_{\rm CaT}(m;t;Z)m^{-\alpha}\,{\rm dm}}\over
        {\int\limits_{m_l}^{m_u}L_{8600}(m;t;Z)m^{-\alpha}\,{\rm dm}}}
$$
where $L_{\rm CaT}$ is the strength of the CaT absorption feature 
and $L_{8600}$ is the underlying continuum luminosity per unit \AA\  at 
the rest wavelength of 8600\AA. The initial mass function (IMF) 
slope $\alpha=2.35$ (Salpeter 1955) is chosen with mass limits $m_l=1$ 
and $m_u=100$ in solar units. The CO index is computed by a similar 
formula with $L_{\rm CO}$ replacing $L_{\rm CaT}$ and the corresponding 
continuum wavelength being 2.36$\mu$m. It is expressed in magnitude units 
following the conventional definition. Stellar evolutionary models are 
used to obtain the spectral type and luminosity class for a star of 
mass $m$ at evolutionary stage $t$ and metallicity $Z$. Theoretical model 
spectra are then assigned to each of these spectral types to enable us to 
compute the continuum luminosities at 8600\AA\ and 2.36$\mu$m. The spectra 
from Kurucz (1992) are used for effective temperatures of stars above 3500 K. 
At lower temperatures we use the library of cool star spectra compiled 
recently by Lejeune, Cuisinier \& Buser (1996). 

We use stellar evolutionary models from Geneva for starburst evolutions at 
metallicities $Z=0.001$, 0.008, 0.02 and 0.04. These metallicities
encompass a wide range of starburst systems from the metal poor blue 
compact galaxies to the metal rich nuclear starbursts. Geneva models 
referred as standard (Schaller et al. 1992; Schaerer et al. 1993a,b)
assume moderate overshooting of the convective core, and use 
empirical relations for expressing mass loss rate (\mdot) for different 
masses at different metallicities. There are known to be difficulties in 
explaining all the observables with one set of assumptions (Maeder 1994; 
Langer \& Maeder 1995). Since our interest here is to compute quantities 
related to RSGs, we looked for those tracks which correctly reproduce 
the observed ratios of blue to red supergiants (B/R). 
Meynet (1992) has shown that the standard tracks reproduce the observed B/R 
at solar metallicity, while at lower metallicity, models
with mass loss rates enhanced by a factor 2.2 reproduce the observed B/R
better. Hence we used standard tracks at $Z\ge$\zsun\ and 
Meynet et al. (1994) models with enhanced mass loss rates (by a factor
of two over standard \mdot) at lower metallicities.

\subsection{Stellar data for \caii\ triplet and CO index}
 
The most comprehensive investigation on the dependence of the equivalent 
width of the CaT with spectral type, luminosity class and metallicity of 
stars was carried out by Diaz, Terlevich \& Terlevich (1989) (DTT hence 
forth). They defined the CaT equivalent width, EW(CaT), as the sum of 
the equivalent widths of two of the brightest lines ($\lambda\lambda$ 8542, 
8662 \AA) and provided an expression relating EW(CaT) with $\log g$ 
and $\log (Z/$\zsun). For metal rich stars ($Z\ge$\zsun) the EW(CaT) 
increases with decreasing surface gravity and for supergiants, it increases 
with increasing metallicity. The dependence on effective temperature was 
found to be negligible for their sample of cool stars. Independent 
observational studies by Zhou (1991) and Mallik (1994) support the 
above findings. Theoretical computations of Jorgensen, Carlsson \& 
Johnson (1992) (JCJ henceforth) successfully reproduced all the basic 
trends established by DTT. However they found the theoretically based 
relations involving higher order terms in $\log g$ are better fits
to the observed data than the linear relation of DTT, which
underestimates the EW(CaT) by $\sim 2$--5\AA\ in metal rich red supergiants. 
Hence we use JCJ relation for $Z\ge$\zsun\ and DTT relation at lower
metallicities.

Data on the CO band is not as complete as that for the CaT. The best
published data-set so far is that of Kleinmann \& Hall (1986) for a sample
of 26 stars covering a wide range of spectral type and luminosity classes, 
but at solar metallicity only. Doyon, Joseph \& Wright (1994) defined the
CO index based on the spectra in the above sample and obtained relations 
for the CO index as a function of effective temperature for dwarfs, giants 
and supergiants separately. These make up the basis for our computations. 
The above defined index (CO$_{\rm sp}$) is related to the photometrically
measured index (CO$_{\rm ph}$; Frogel et al. 1978) by 
CO$_{\rm sp} = $1.46 CO$_{\rm ph} - 0.02$.

\section{Synthetic results for \caii\ triplet and CO index}

The evolutionary behavior of the EW(CaT) at the four metallicities is 
shown in Fig.~1a. The following important stages can be recognized during the
evolution of solar metallicity starbursts. \\
Stage 1. EW(CaT) rises to detectable levels in starbursts, 5~Myr after the 
onset of the IB, reaching the highest values of $\sim$11\AA\ between 7--14~Myr. 
\\
Stage 2. The EW abruptly drops to around 5\AA\ at $\sim$15~Myr, remaining at 
these low values up to around 20~Myr. \\
Stage 3. The EW rises again above 6.5\AA\ at 25~Myr, showing
a secondary peak at 60~Myr. \\ 
Stage 4. The EW attains an asymptotic value of around 3\AA\ after 100~Myr.

\begin{figure}[ht]
\centerline{\psfig{figure=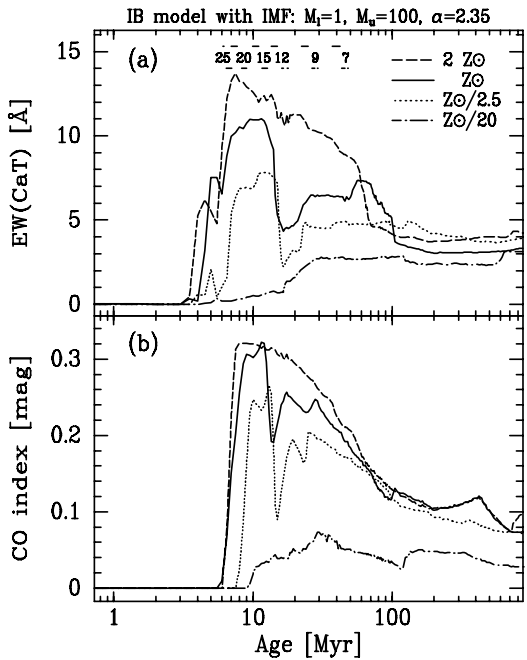,height=9.5cm}}
\caption{
Evolution of (a) the equivalent width of \caii\ triplet and (b) the CO index 
for an instantaneous burst of star formation for the assumed Salpeter 
initial mass function, at the four chosen metallicities. Supergiant 
phases of massive stars at 2\zsun\ (top) and \zsun\ metallicities
are shown by the parallel horizontal bars, with the numbers in the
middle denoting the stellar masses in \msun.
}
\end{figure}
The evolutionary behavior at $Z=$\zsun/2.5 is qualitatively similar to
\zsun\ models. However the peak values, as well as the contrast between
Stages 1,3 and 4 above, are lower. The trend of decreasing
contrast continues at lower metallicity resulting in steady values of
around 2\AA\ throughout the Stages 3 and 4 for \zsun/20 model.
The most significant feature during the evolution of low metallicity
starbursts is the absence of the primary peak at 10~Myr (Stage 1).
At intermediate metallicities (\zsun/5; not plotted), the 10-Myr peak is 
present (2\AA) but at values lower than at Stage 3 (5\AA).
At higher than solar metallicities, Stage 2 is missing resulting in a
smooth decrease of the EW from Stage 1 to the end of Stage 3.

The Stages 1--4 as well as their dependence on metallicity can be understood 
by identifying the individual phases, which contribute significantly to the 
EW(CaT), during stellar evolution. The cool giant and supergiant 
phases of different stellar masses are depicted in Fig.~1 by horizontal 
bars at the upper part of the diagram. The primary peak at $\sim10$~Myr at 
all the metallicities is directly related to the cool RSG population 
of progenitor masses between 15--25\msun. The supergiants at the lowest 
metallicity we have considered, spend most of their post-main sequence 
evolution as blue supergiants, and hence fail to show the peak at 10 Myr. 

Note that the horizontal bars depicting the cool giant and supergiant phases 
have a discontinuity for stars of mass $\le12$\msun\ at solar metallicity. 
This break is due to the qualitatively different post-main sequence evolution 
of 12\msun\ and lower mass stars for $Z\le$\zsun. These stars go through the 
well-established ``blue loop'' in the Hertzsprung-Russel diagram during 
the core helium burning phase, 
returning to the redder phase only at the end of core helium burning. Thus, 
there is a sudden reduction in the number of RSGs in the starburst,
which explains the Stage 2. ``Blue loops'' during core helium burning is
a common property of stellar evolutionary models for $Z\le$\zsun, 
and hence \zsun/2.5 model behaves qualitatively similar to the \zsun\ model.
On the other hand, ``blue loops'' are absent for higher metallicity models
and hence there is a smooth transition from the RSG-dominated phase to 
the red giant and asymptotic giant branch dominated phases at $Z=2$\zsun. 

Successively lower mass stars spend a lesser fraction of their post-main 
sequence lifetime in ``blue loops'', resulting in the accumulation of red
giant and asymptotic giant branch stars, leading to the secondary peak at
60~Myr for \zsun\ models. The adopted scaling between mass-loss rate and 
metallicity and also the corresponding phases for a star of given mass 
being warmer at lower metallicities result in the net decrease of the
number of stars contributing to CaT absorption at lower metallicities. 
This is the reason for lower peak values as well as lesser contrast
between different Stages at lower metallicities.
After 100~Myr, the EW(CaT) is controlled by the cool giant and dwarf 
stars of mass $<5$\msun. 

Evolution of the \cosp\ index is shown in Fig.~1b. Note that the
metallicity dependence in this plot arises only due to the dependence of 
stellar evolution on metallicity; there is no data-set on its 
dependence on line formation. The evolutionary behavior, as well as its
dependence on metallicity, can be understood by the same physical
phenomenon as that for the EW(CaT). However the following quantitative 
differences in the four Stages are noteworthy.\\
Stage 1. The RSG peak is narrower for $Z\le$\zsun\ (8--12~Myr).\\
Stage 2. The CO index rises faster after the drop following the ``blue loop''
phase.\\
Stage 3. The secondary maximum occurs at 20--30~Myr, decreasing smoothly then 
onwards up to around 100~Myr. \\
Stage 4. The CO index do not reach asymptotic values beyond 100~Myr.

These differences arise firstly, due to a stronger dependence of the CO index 
on the effective temperature and secondly, due to the same stars contributing 
to the continuum 
as well as the CO absorption strength. The CO index of the starburst is 
essentially that of the most luminous cool star, as these cool stars 
contribute both to the line depth as well as the continuum. In contrast 
the continuum at CaT has a significant contribution from hotter stars and 
hence the EW(CaT) evolves differently than that of the CO index. 

\section{Comparison with LMC clusters}

The blue clusters of the LMC provide a unique opportunity
of checking the evolutionary results discussed above, before these results
can be used to derive ages of starburst systems. The special advantage
of these clusters is the availability of well-determined turn-off ages. 
As discussed earlier, Bica et al. (1990) had obtained the EW(CaT) for 
each cluster, treating each cluster as a single entity. 
Bica et al. also compiled the CO indices and broad band colors for each
of these clusters, which are ideal for comparison with our results.
We superimpose our model results on these observed data in Fig.~2. 
The strength of the fainter $\lambda$8498\AA\ line has been subtracted from
the Bica et al. value to covert the observed values to the more commonly
used definition of DTT. Photometrically derived CO indices are converted 
into \cosp\ using the relation given in sec.~2.1. 
The available $J-K$ and $B-V$ colors are also plotted in the last two panels.
The solid and dashed lines correspond to models using enhanced and 
standard mass loss rates respectively, both at \zsun/2.5 metallicity. 

\begin{figure}[ht]
\centerline{\psfig{figure=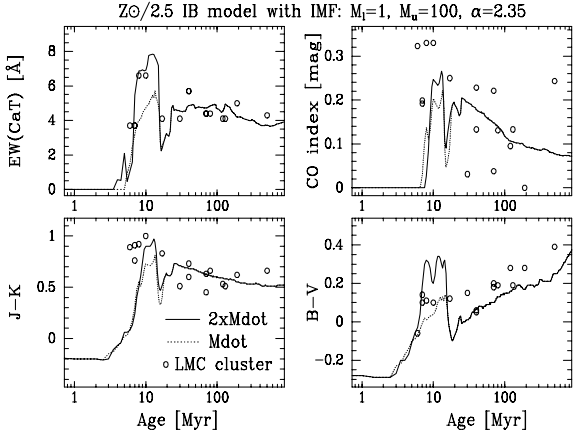,height=6cm}}
\caption{
Comparison of the computed quantities at $Z=$\zsun/2.5 with the observational 
data of Bica et al. (1990) for the LMC clusters. 
Stellar evolutionary models incorporating twice the standard 
mass-loss rate (solid line) reproduce the observed characteristics better.
}
\end{figure}
The ages of the clusters with the highest EW(CaT), CO index and $J-K$ color
agree very well with the RSG dominated bump (Stage 1) in our model.
The absolute values of the RSG peak with stellar evolution incorporating 
standard mass loss rate is too low to account for the observations,
confirming the under-estimation of RSGs in these models.
Usage of enhanced mass loss rate models correctly reproduces the EW(CaT)
and $J-K$ color. The observed CO index is still higher by around 0.08~mag, 
which might be caused by errors introduced in converting the 
photometrically observed indices to spectroscopic indices.
The larger scatter in this plot compared to the other plots also 
suggests larger errors in the observed CO indices. Thus we conclude 
that the stellar evolution with enhanced mass-loss rates, and not the 
standard, is a better representation of the observed values in LMC. 

Older clusters have EW(CaT), CO index and $J-K$ as predicted for 
Stage 3 in our model. However only the lower boundary 
of the observed $B-V$ colors agree with the model predictions
for Stage 3 with the majority of the clusters having colors 
around 0.1~magnitude redder. A similar trend is also seen in $U-B$ 
colors (not plotted), which is most likely due to the internal 
reddening in the clusters. It is interesting to note that the 
observed $B-V$ (and also $U-B$) colors during the RSG phase are bluer 
than the predictions of the enhanced mass loss rate models. 

\section{Discussions and Outlook}

Having studied the evolutionary behavior of the CaT equivalent width 
and the CO index, we now discuss their usage in estimation of ages of 
starburst regions from observed data. Diagrams involving dimensionless 
quantities play an important role in this direction. Most of such 
attempts hitherto are based on the presence of massive O stars or 
Wolf-Rayet stars and hence are applicable only as long as the 
starburst nebula is traceable. The end of nebular phase coincides with the
beginning of the RSG phase for an IB, as the ionization is mostly 
controlled by stars having masses higher than the progenitors of the
RSGs ($\sim30$\msun). This is illustrated in the top panel of Fig.~3. 
It can be seen that by the time the EW(CaT) attains the peak value the
\ha\ emission equivalent width has dropped below the detection limit.
We investigated various diagrams involving EW(CaT), CO index and 
colors for the four metallicities, and found the EW(CaT)--CO index plane 
to be the best suited for the purpose of age determination.
The loci of IB and continuous star formation (CSF) models 
at $Z=2$\zsun\ and \zsun\ are shown in Fig.~3. 
Selected ages in Myr are indicated along the locus in this plot. 

\begin{figure}[ht]
\centerline{\psfig{figure=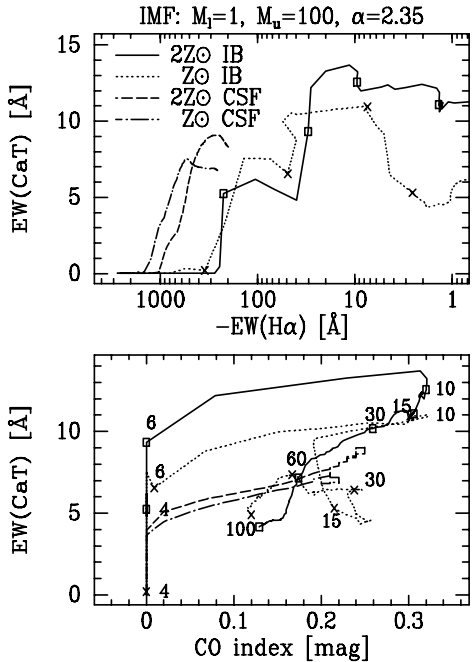,height=9.5cm}}
\caption{
Dimensionless diagrams involving observational quantities for
$Z=$\zsun\ and $Z=$2\zsun\ models with instantaneous burst and 
continuous star formation (CSF) at a constant rate.
The numbers next to tick marks are starburst ages in Myr. For the CSF
model values stabilize at $\sim$15~Myr.
}
\end{figure}
In the EW(CaT) vs EW(\ha) plot, the IB and CSF models follow quite different 
tracks and hence can be most effectively used to infer the history
of star formation; e.g. regions with detectable EW(CaT) and emission EW(\ha)
greater than 300\AA\ can be described only by CSF. For CSF models 
asymptotic values are reached in $\sim$15~Myr. Once the history of star
formation is known, the EW(CaT) vs CO index plot can be used to derive 
accurate ages of starburst regions. The ambiguity in the estimated age 
increases with decreasing metallicity for IB models, 
with the result that similar plots for $Z<$\zsun\ have limited use.
However our models give a definite upper limit for low metallicity systems.
The blue compact galaxies with $Z<$\zsun/5 are not expected 
to have EW(CaT) and CO index values above 5\AA\ and 0.1~dex respectively.
These values are insensitive to mass limits as long as $m_u\ge30$\msun\ 
and $m_l\le3$\msun, and are only weakly sensitive to the IMF slope. 

Before we end, we point out the limitations of the present work and discuss
where further improvements are necessary. Our main purpose in this work was to
investigate the important stages during evolution of EW(CaT) and CO index in 
starburst regions and their metallicity dependence, and accordingly we
kept the starburst model at the simplest level. Observed values in
nuclear starbursts may have some contribution from the bulge light (see e.g.
Garc\'{\i}a-Vargas et al. 1997), and hence usage of Fig.~3 for nuclear
starbursts may require a separate treatment of the bulge. However Fig.~3
is directly applicable for high metallicity giant \hii\ regions in disks 
of galaxies. Our next step is to study the dilution of the computed
quantities as a function of fractional contribution of the bulge light,
which will enable direct comparison with observed data of starburst nuclei.

\acknowledgements{It is a pleasure to thank Enrique Perez for his keen
interest throughout this project and also for a critical reading of the
manuscript. I acknowledge the support from CONACYT through the grant, 
Catedra Patrimonial Nivel II (Ref.:960028).
}


\begin{references}

\reference{} Bica, E., Alloin, D., \& Santos Jr., J.F.C. 1990, A\&A, 235, 103

\reference{} Charlot, S., \& Bruzual, A. 1991, ApJ, 367, 126

\reference{} Charlot, S. 1996, in From Stars to Galaxies, ASP Conf. Ser. 98, 
            275 (Eds. C. Leitherer, U. Fritze-v. Alvensleben, \& J. Huchra) 

\reference{} Diaz, A.I., Terlevich, E., \& Terlevich, R. 1989, MNRAS, 239, 325

\reference{} Doyon, R., Joseph, R.D., \& Wright, G.S. 1994, ApJ, 421, 101 

\reference{} Frogel, J.A., Persson, S.E., Aaronson, M., \& Matthews, K. 1978, 
             ApJ, 220, 75  

\reference{} Garc\'{\i}a-Vargas, M.L., Gonzalez, R.M., Perez, E., et al. 1997
              ApJ (in press)
 
\reference{} Jorgensen, U.G., Carlsson, M., \& Johnson, H.R. 1992, 
             A\&A, 254, 258

\reference{} Kleinmann, S.G., \& Hall, D.N.B. 1986, ApJS, 62, 501

\reference{} Kurucz, R.L. 1992, in Stellar Populations of Galaxies, IAU Symp. 
         149, 225 (Eds B.~Barbuy \& A.~Renzini: Kluwer, New York)

\reference {} Langer, N., \& Maeder, A. 1995, A\&A, 295, 685

\reference {} Leitherer, C., Alloin, D. Fritze-v. Alvensben, et al. 1996,
               PASP, 108, 996 

\reference {} Lejeune, T., Cuisinier, F., \& Buser, R. 1996, in 
             From Stars to Galaxies, ASP Conf. Ser. 98, 94
             (Eds. C. Leitherer, U. Fritze-v. Alvensleben, \& J. Huchra)

\reference{} Maeder, A. 1991, A\&A, 242, 93

\reference{} Maeder, A. 1994, Space Science Reviews, 66, 349

\reference{} Mallik, S.V. 1994, A\&AS, 103, 279

\reference{} Mayya, Y.D. 1995, AJ, 109, 2503

\reference{} Meynet, G. 1992, in ``The feedback of chemical evolution on
     stellar content of galaxies'', Eds. D. Alloin \& G.Skasinska, Obs. Paris,
     p. 40

\reference{} Meynet, G., Maeder, A., Schaller, G., Schaerer, D. \&
           Charbonnel, C. 1994, A\&AS, 103, 97

\reference{} Salpeter, E.E. 1955, ApJ, 121, 161 

\reference{} Schaerer, D., Meynet, G., Maeder, A., 
           \& Schaller, G. 1993a, A\&AS, 98, 523

\reference{} Schaerer, D., Charbonnel, C., Meynet, G., Maeder, A., 
           \& Schaller, G. 1993b, A\&AS, 102, 339

\reference{} Schaller, G., Schaerer, D., Meynet, G., \& Maeder, A. 1992, 
           A\&AS, 96, 269

\reference{} Terlevich, E., Diaz, A.I, Pastoriza, M.G., Terlevich, R. \& 
           Dottori, H. 1990, MNRAS, 242, 48p

\reference{} Zhou, X. 1991, A\&A, 248, 367

\end{references}
\end{document}